\documentclass[lettersize,journal]{IEEEtran}
\usepackage{amsmath,amsfonts}
\usepackage{algorithmic}
\usepackage{algorithm}
\usepackage{array}
\usepackage[caption=false,font=normalsize,labelfont=sf,textfont=sf]{subfig}
\usepackage{textcomp}
\usepackage{stfloats}
\usepackage{url}
\usepackage{verbatim}
\usepackage{graphicx}
\usepackage{cite}
\usepackage{colortbl} 
\usepackage[table]{xcolor} 
\usepackage{xcolor}
\begin{document}


\title{A UAV-assisted Wireless Localization Challenge on AERPAW}


\author{
Paul Kudyba, Jaya Sravani Mandapaka, Weijie Wang, 
Logan McCorkendale, Zachary McCorkendale, Mathias Kidane, 
Haijian Sun, Eric Adams, Kamesh Namuduri, Fraida Fund,
Mihail Sichitiu, Ozgur Ozdemir \\

\thanks{P. Kudyba and H. Sun are with School of Electrical and Computer Engineering, University of Georgia, Athens, GA, USA.  \\
J. S. Mandapaka, L. McCorkendale, Z. McCorkendale, M. Kidane, and K. Namuduri are with Department of Electrical Engineering, University of North Texas, Denton, TX, USA. \\
W. Wang and F. Fund are with epartment of Electrical and Computer Engineering, NYU Tandon School of Engineering, Brooklyn, NY, USA. \\
E. Adam is with Delmont Systems LLC, Hurst, TX, USA.  \\
M. Sichitiu and O. Ozdemir are with Department of Electrical and Computer Engineering, North Carolina State University, Raleigh, NC, USA. }
}
\maketitle

\maketitle

\begin{abstract} 
As wireless researchers are tasked to enable wireless communication as infrastructure in more dynamic aerial settings, there is a growing need for large-scale experimental platforms that provide realistic, reproducible, and reliable experimental validation. To bridge the research-to-implementation gap, the Aerial Experimentation and Research Platform for Advanced Wireless (AERPAW) offers open-source tools, reference experiments, and hardware to facilitate and evaluate the development of wireless research in controlled digital twin environments and live testbed flights.
The inaugural AERPAW Challenge, ``Find a Rover,'' was issued to spark collaborative efforts and test the platform's capabilities. The task involved localizing a narrowband wireless signal, with teams given ten minutes to find the "rover" within a twenty-acre area. By engaging in this exercise, researchers can validate the platform's value as a tool for innovation in wireless communications research within aerial robotics. This paper recounts the methods and experiences of the top three teams in automating and rapidly locating a wireless signal by automating and controlling an aerial drone in a realistic testbed scenario.
\end{abstract}


\section{Introduction} 
\IEEEPARstart{L}{ocalizing} 
wireless radio transmissions has numerous use cases ranging from search and rescue operations, wildlife tracking, to finding jammers (intentional or not), as well as tracking intruding Unmanned Aerial Vehicles (UAVs)~\cite{kwonRFSignalSource2023a}. Many methods of localizing wireless transmitters are common, including using Radio Frequency (RF)  sensors at fixed locations, using vehicles or manned aircraft with RF receivers, or searching on foot (commonly referred to as ``fox hunting''). Among these methods, using UAVs for localization of RF sources has the potential to alleviate many of the drawbacks of the other methods, resulting in a cost-effective solution, yet quickly covering a large geographical area and with the potential for very accurate results.

Since optimal solutions are computationally intractable, the methods for wireless localization in the research literature often try to strike a balance between computational complexity and their ability to handle noisy measurements and heterogeneous environments; in the end, the proof is in the pudding, and an experimental approach to performance evaluation is often the best indication on the performance of an approach in the real world for this type of research. A significant hurdle for experimentation with UAVs and wireless communications equipment is the considerable effort involved in building the UAVs and their payloads, obtaining RF permits (such as FCC approvals), and the availability of FAA safety pilots for the UAVs.

The Aerial Experimentation and Research Platform for Advanced Wireless (AERPAW)~\cite{AERPAW} has been designed and built to facilitate the development and testing of this type of research. In the AERPAW platform, researchers can develop their experiments involving wireless communications, UAVs, and Unmanned Ground Vehicles (UGVs) in a digital twin emulation environment and then transfer their experiments to the physical testbed. The experiments are then executed using real UAVs, UGVs, and radio transceivers, and the results are subsequently returned to the researchers in the emulator, where a new iteration can take place. In the Fall of 2023, AERPAW organized a student competition named AERPAW Find a Rover (AFAR) that challenged teams of students to program a drone equipped with a wireless receiver to find the location of a hidden transmitter located on a UGV on the ground. This paper details the approaches taken by the top three finalists as well as lessons learned about the AERPAW platform in particular and digital twins in general.

The remainder of the paper is organized as follows. Section~\ref{sec:AFAR} and \ref{sec:challenge} present details about the AERPAW platform and the AFAR challenge. Section~\ref{sec:methodology} describes the approaches taken by each of the three top teams in the challenge. Section~\ref{sec:results} shows and comments on the results of the experiments in the field and compares them with the results from the digital twin. Section~\ref{sec:conclusions} concludes the paper.



\begin{figure}
    \centering
    \includegraphics[width=1\linewidth]{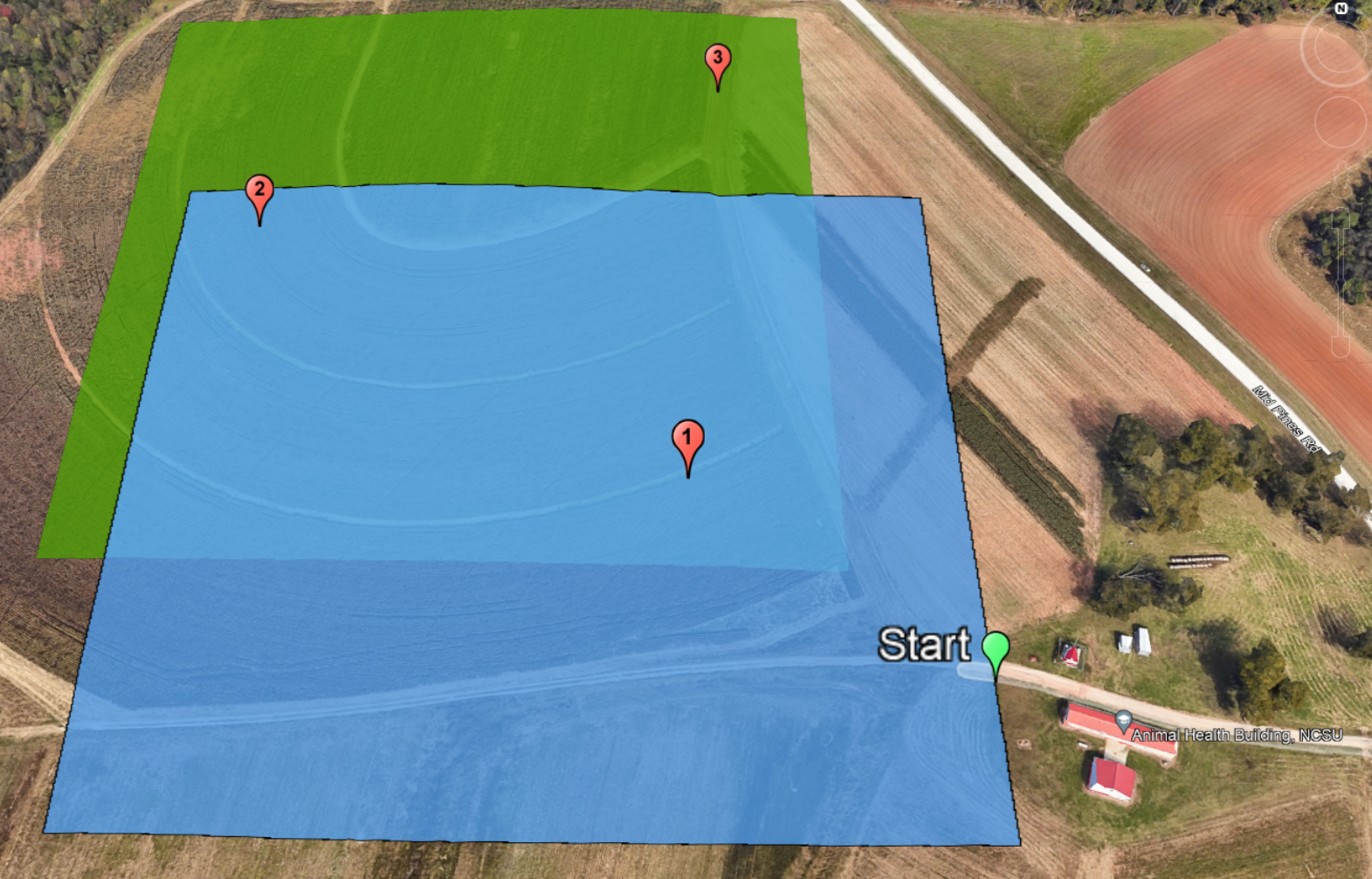}
    \caption{Physical setup of the AFAR competition, with the blue area showing the allowable locations for the UAV, the green area the hiding area of the rover, and the three markers showing the three hiding spots during the competition.}
    \label{fig:aerpaw_locations}
\end{figure}

\subsection{The AERPAW Platform} \label{sec:AFAR}

AERPAW is the third of the original four Platforms for Advanced Wireless Research (PAWR), which are a set of experimental platforms sponsored by the National Science Foundation (NSF) in partnership with an industry consortium. The four platforms allow wireless researchers from industry and academia to perform wireless experiments at scale, in a real outdoor environment. While all four platforms enable advanced wireless research, only AERPAW enables controlled mobility, by allowing the researchers to program autonomous aerial and ground vehicles.

At its core, AERPAW consists of a physical testbed and its supporting facilities. The physical testbed consists of fixed and mobile nodes. The fixed nodes comprise eight 20m tall fixed towers (or pole/roof-top mounts) with an enclosure at the base and radio equipment installed at each location. The fixed nodes, in addition to a common set of USRPs (Universal Software Radio Peripheral), feature a heterogeneous collection of radio equipment that includes a 4G/5G NSA Ericsson cellular network, RF sensors, and LoRa Gateways. The mobile nodes are comprised of a vehicle and a portable node. In general, (with one exception) any portable node can be mounted on any vehicle depending on the experiment's needs. The vehicles relevant for the AFAR challenge are a Large AERPAW Multicopter (LAM), and the first AERPAW rover. Both vehicles can handle the Large Portable Nodes (LPN), and during the experiments in the AFAR challenge, each vehicle carried one such node.

The LAM is a capable UAV designed by the AERPAW team to carry the 3-4~kg of a typical LPN for up to 40~minutes. The maximum payload of the LAM is limited by FAA regulations to 13~kg (with a corresponding reduction in endurance). All the components of an LPN are centered around the USRP (for this competition, a B205 mini was used): an Intel~NUC~10 (Intel i7-5550U with 64~GB~RAM and 1~TB~SSD) is used as a companion computer, 
%
and a 1~W wide-band low-noise power amplifier with filters is used at the front ends of the USRP.
Power for the LPN is provided by the vehicle batteries. Each LAM is controlled by an autopilot that is capable of executing the commands sent by the companion computer of the attached LPN. The operating frequency during the AFAR challenge was 3.4~GHz.

An essential supporting system for the AERPAW physical testbed is its digital twin. The AERPAW digital twin environment is designed to allow users to develop experiments without requiring direct access to the AERPAW UAVs and radios. While other testbeds allow their users direct access to the physical resources of the testbed, due to the inclusion of programmable UAVs in the testbed, AERPAW users have to use the digital twin to develop their experiments. In the digital twin, three hardware elements are virtualized: the USRPs, the channel propagation, and the UAVs. Standing in for the USRPs in the testbed, the AERPAW team has developed virtual USRPs that can be discovered by the USRP Hardware Driver (UHD) and used like hardware-based USRPs by any software that uses UHD. The AERPAW channel emulator is designed to forward (after considering fading and channel impairments) the signals between the virtual USRPs of all participating nodes in an experiment. Finally, a software-in-the-loop setup is used to emulate the drones. The drone emulator is periodically updating the wireless channel emulator with the positions and orientations of the virtual drones such that the channel emulator can take into account the relative positions of all virtual USRPs in the testbed, including their relative antenna gains. Once an experiment is developed in the digital twin, it can then be transferred to the testbed (by moving docker containers) and executed in batch mode; subsequently, the experiment results are returned to the experimenters in the digital twin.

When comparing results from the testbed and digital twin emulation, the emulation of the drones is quite accurate: both the real as well as the virtual drones have identical autopilot software, and both the real and the virtual drones are more than capable of executing the commands given by the autopilots (i.e., their performance limits are far in excess of the demands of the autopilot). On the other hand, the wireless channel emulator was primarily designed to enable the development of experiments that will be executed on the testbed to obtain the results. As such, the accuracy of the channel emulator is nowhere close to the accuracy of the drone emulator. In short, the channel emulator at the time of the challenge used a free-space propagation model with 10~dB of added white noise, while the propagation in the physical testbed was considerably noisier (with up to 30-40~dB of noise at times).
In Section~\ref{sec:results}, it will be clear that this discrepancy between the digital twin propagation model and the testbed wireless propagation has had a clear detrimental effect on the teams that tuned their approach to the propagation characteristics of the digital twin (while the evaluation of the results was performed in the real testbed).

\subsection{The AFAR Competition} \label{sec:challenge}

In the summer of 2023, the AERPAW team decided to hold a student competition on the AERPAW platform for several reasons: first, a competition will raise the visibility of the platform among wireless researchers, which could convert into users at a later time; furthermore, a competition will exercise and validate the entire platform design, and allow the AERPAW team to identify any significant obstacles in using the platform. Finally, the problem considered has many real-life use cases, and a competition in a real environment may provide interesting insights toward solving the problem.

The problem considered in the AFAR challenge is as follows: the organizers (AERPAW)  {\em hide} a rover at an unknown location while the rover is continuously transmitting a narrowband radio signal, while the competitors program a drone equipped with a wireless receiver in order to {\em find} the rover (i.e., estimate its location). While the problem is {\em theoretically} relatively simple, noisy measurements (from the reduced bandwidth and hardware impairment) considerably complicate data processing as well as the design of the search algorithms.

Both the hidden rover and the drone feature wideband {\em nearly} omnidirectional antennas, which result in a local minimum when the drone is directly above the rover. Furthermore, during the drone movement, the drone antenna tilts up to five degrees in the direction of the movement, introducing further uncertainty in the measurements. While the {\em maximum} signal strength decreases reliably with the distance between transmitter and receiver, signal strength drops of 30-40~dB were very common throughout the experiment.

Figure~\ref{fig:aerpaw_locations} shows the physical setup of the competition: the blue rectangle shows where the UAV is allowed to fly (at a minimum altitude of 20~m and maximum of 110~m). The green rectangle shows the allowed locations for the rover. For reference, the sides of the rectangles were between 270~m and 300~m. The competitors were allowed to fly the UAV in any orientation they desired and at any speed under 10~m/s.

The three red markers show where the rover was actually hidden by the organizers during the competition. Naturally, the competitors were not aware of these locations before the competition. The three ``hiding spots'' for the rover were intended to be increasingly difficult, with the first one close to the start location of the drone, the second one farther away (and close to an edge of the allowable area), and the third one intentionally placed in a region where the drone was not allowed to travel to. The same three locations were used for all runs for all competitors. The competitors did not have access to the rover in any way during the competition.

To balance localization speed and accuracy, two separate estimates were required from the teams (during the same flight): the first estimate is a {\em fast} estimate, which is required after three minutes from the start of the search. The second estimate is the {\em final} estimate, which is required after ten minutes from the start of the search. Separate scores and awards were made for each of the two estimates.

In preparation for the competition, the AERPAW team developed the channel sounder (based on a pseudo-random noise sequence) comprising the transmitter on the rover and the receiver on the UAV. Each measurement consisted of two numbers: a signal strength and a confidence level (which measured how strong the signal was in comparison with the background noise). Furthermore, the AERPAW team also provided sample code showing how to integrate UAV control with radio measurements. The sample code (a simple type of gradient descent) assumed that the channel measurements are monotonically varying with the distance between the rover and the UAV, which is an assumption that does not hold (at all) in the real testbed. All the participating teams significantly changed the sample code (they only used the primitives the AERPAW team provided for moving the UAV and collecting radio measurements). Section~\ref{sec:methodology} provides details on the methodology employed by each of the top three teams.

For fairness, none of the AERPAW students were allowed to participate in the competition, and none of the principal investigators were allowed to mentor any participating teams. The score for each team was computed by computing the average error for their estimates for either the fast or the final estimates (or both). The team with the lowest average errors wins (in each category).

\section{Design Methodology and Experiences} 
\label{sec:methodology}


\subsection{New York University (NYU) Team}
The NYU team used a reference experiment provided by AERPAW as a starting point. In this baseline solution, the UAV measures the received signal strength at fixed intervals. If the signal strength decreases between the beginning and end of an interval, the UAV turns 90~degrees. We used the baseline to identify areas of improvement, and then designed our final solution - based on Bayesian optimization - to address these.

In initial experiments with the baseline solution, we identified these challenges:

\begin{itemize}
    \item With a fixed interval size, the UAV either overshoots the rover location (too-large interval) or moves very slowly (too-small interval). At the end of the flight - when the UAV is close to the rover - the UAV flies in a rectangle around the rover, and cannot get closer.
    \item Early in the search, ``decreasing signal strength'' is a relatively reliable indicator of whether the UAV is moving toward the rover or away. Later in the search, however, when the UAV is already close to the rover, this is no longer a reliable indicator because of the noisy relationship between signal strength or distance.
\end{itemize}

The first challenge was easily addressed with small changes to the baseline solution. The first change, modeled after learning rate annealing in gradient descent, was to gradually reduce the interval during flight time. Then the UAV can move fast at the beginning of the search, and take smaller steps at the end. Next, inspired by the idea of momentum in gradient descent, we added an accumulating additive term to the interval, so the UAV moves faster in the same direction if the signal strength increases in successive intervals. 
The team also tracked boundaries in each of the four cardinal directions, updated each time signal strength decreased while the UAV was moving in that direction, 
and made the UAV more cautious about crossing a ``boundary,'' when it had previously seen signal strength decrease beyond that point.

With the changes to the baseline solution mentioned above, the trajectory of the UAV was much more efficient, and it reached the approximate location of the rover quickly. However, because of the second challenge mentioned above, the location estimate did not improve much with additional flight time.
Consequently, the NYU team switched to an approach based on Bayesian optimization to fit a Gaussian process (GP) regression~\cite{rasmussenGaussianProcessesMachine2006}. The GP regression can learn from noisy data by adding a white kernel to interpret the noise. Also, Bayesian optimization is well suited for optimizing a problem where it is expensive to sample new points (in this case, consumes more flight time) in the parameter space. The algorithm was implemented with the \texttt{bayesian-optimization} package in Python~\cite{bayesopt}, which turned out to be very effective. The team used the digital twin environment in AERPAW to experiment with different kernels for the GP regression, different acquisition functions for the Bayesian optimization, and hyperparameters for both the kernel and acquisition function, in order to further improve the results.

The NYU team also added two features to our solution to accommodate specific elements of the AFAR challenge. First, we need the UAV to have a reasonable fast estimate after three minutes. Therefore, we began each search by traversing the south and west edges of the UAV geofence, to identify the latitude and longitude at which the highest signal strength was observed, and used that as an initial estimate. Second, we had to address the possibility of the rover being outside the UAV geofence. In that case, 
we flew the UAV perpendicular to the boundary, then used a previously fitted linear regression to estimate the distance between the boundary and the rover.

    \begin{figure}[!ht]
        \centering
        \includegraphics[width=1\linewidth]{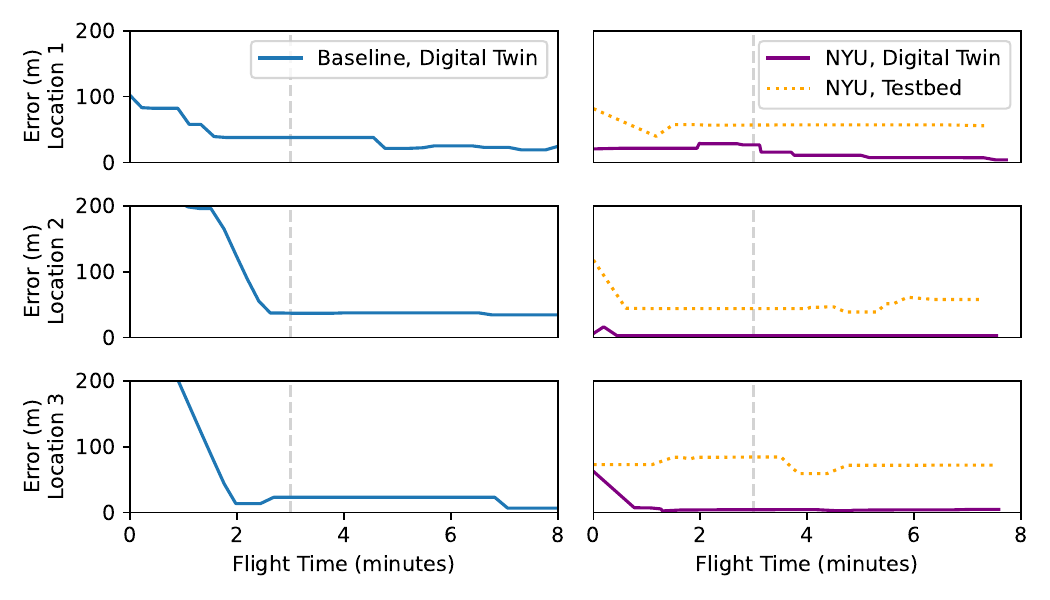} 
        \caption{Error of location estimate of the NYU team solution, and the baseline solution provided by AERPAW. The dashed vertical line marks the three-minute mark, at which the fast estimate is due.} 
        \label{fig:NYUteam} 
    \end{figure}

As shown in Figure~\ref{fig:NYUteam}, the NYU team achieved good improvement over the baseline solution in the digital twin environment, although the location estimate was not as accurate in the physical testbed due to differences in the signal propagation dynamics. We attribute this success to the ability to conduct extensive experiments in the digital twin environment, and to the reference code that allowed the team to start with a working solution right away and then iterate on it. 

\subsection{University of North Texas (UNT) Team}
The UNT team developed a recursive algorithm that can be used to scan and locate the rover. The team's approach was to execute a perimeter sweep of the overlap between the two geofences (rover and UAV), as shown in Figure~\ref{fig:aerpaw_locations}. 
Although the rover was not guaranteed to be located in this region, it was the largest searchable region within which the rover may be located. Its rectangular nature lent itself to an edge traversal, with periodic measurements logged, and the location of the highest power value recorded on a per edge basis. 

    \begin{figure}[!ht] 
        \centering
        \includegraphics[width=0.97\linewidth]{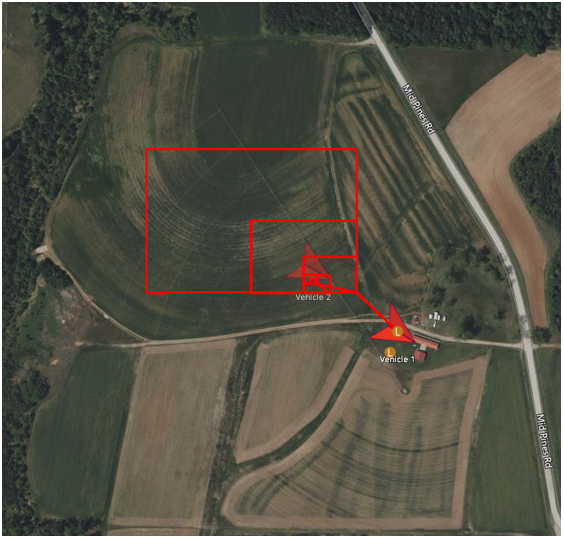} 
        \caption{UNTs recursive algorithm final result shown in QGroundControl, the open-source flight control and mission planner program used by AERPAW.} 
        \label{fig:UNTAlgorthim} 
    \end{figure}
    
As the drone continued along its path, it would sample a signal power value every 0.2 seconds. This signal was then pushed to a buffer, and when the buffer reached a length of 8 recorded samples, the average was calculated and stored. The average signal power values calculated are then associated with the center location of the buffer. This allowed for a continuous 1.6-second interval scan of the signal power strength along the perimeter. All the signal power strengths and their locations were then stored for further use. Four positions were then selected, one on each side of the rectangle determined by the greatest signal power strength. 

Implementing the sampling buffer in our approach was to account for external noise from other unknown sources. The sampling and averaging of the radio frequency power values negated the effects that random noise spikes would have had during the search. For example, if the drone traverses a side of the perimeter and experiences a sudden burst of radio frequency noise and records it, that position would be the point of highest recorded value, resulting in an inaccurate measurement. With the sampling and averaging process, we can filter out all the instances of noise and obtain a more refined and accurate search area. 

Once a full perimeter sweep has concluded, the measured signal values are evaluated to determine the location of the highest recorded power values on each side of the rectangle. The four locations are then used to create line segments from points on opposite sides of the rectangle. Vertical and horizontal line segments are created using Pyturf's Geojson~\cite{PyturfPyturf2024} format by inserting two locations (with both latitude and longitude). These lines were then used in Pyturf's line intersect function to calculate the location at which the two lines intersect, predicting the initial guess. The recursive algorithm then repeats the perimeter sweep by selecting a quadrant that contains the previous intersection location. After recording the greatest received power strengths and locations, the location associated with the overall greatest received power strength was compared to all the current corner locations of the perimeter sweep to determine the next quadrant. 

The closest corner location to the highest received signal power location is then used to select the next quadrant to search. This corner is determined to be the first corner for the next recursion; the remaining three are derived by leveraging their relationship to the initial corners. Two of the three corner points are derived using the half-length of each side segment to determine the point location. The final point is determined by creating two line segments to find the center position of the current search perimeter. Now that the corners of the quadrant are defined, the UAV calculates the distance between all the corner locations, where the shortest distance determines the starting location. The corner locations are updated for the next recursive perimeter sweep and are used to define the path the UAV will traverse. Upon the last recursion, the intersection point is selected as our final and best guess for the location of the missing rover, as shown in Figure~\ref{fig:UNTAlgorthim}.

One of the most notable experiences was learning the difference between the emulation environment and the real-world execution of the experiment. The UNT team took into consideration the potential interference of signal noise from unknown sources that were not present in the emulated environment. This reality was not stated in the competition description and rule book, but upon reviewing the signal values in the emulation, the team realized the signals were too clean and did not show similar instances of simulated noise. Once this was realized, the search algorithm was adapted to account for noise, which was the sampling buffer and averaging of the recorded values. This allowed for an advantage in real-world flight. 

\subsection{University of Georgia (UGA) Team}

The UGA approach centered on achieving rover localization by expeditiously solving for the unknown path loss through continuously updating a 2D regression function or radio map~\cite{wilsonEfficientlySamplingFunctions2020, santosMultirobotLearningCoverage2021}. The algorithm autonomously navigates the UAV to estimate the transmitter’s position from a handful of strategically collected radio samples. Due to the limited flight time and access to real data, the regression model needed to make minimal assumptions~\cite{finkOnlineMethodsRadio2010}. Therefore, the team chose GP regression for its sparse, adaptable, and statistically rigorous path loss estimation~\cite{shresthaRadioMapEstimation2023, deisenrothGaussianProcesses, duvenaudAutomaticModelConstruction, matthewsGaussianProcessBehaviour2018a}.
The GP radio map was updated with new valid samples, and Bayesian optimization guided the UAV to optimal sample locations to improve the radio map~\cite{wangRecentAdvancesBayesian2023}.

    \begin{figure*}[!ht] 
        \centering
        \includegraphics[width=0.97\linewidth]{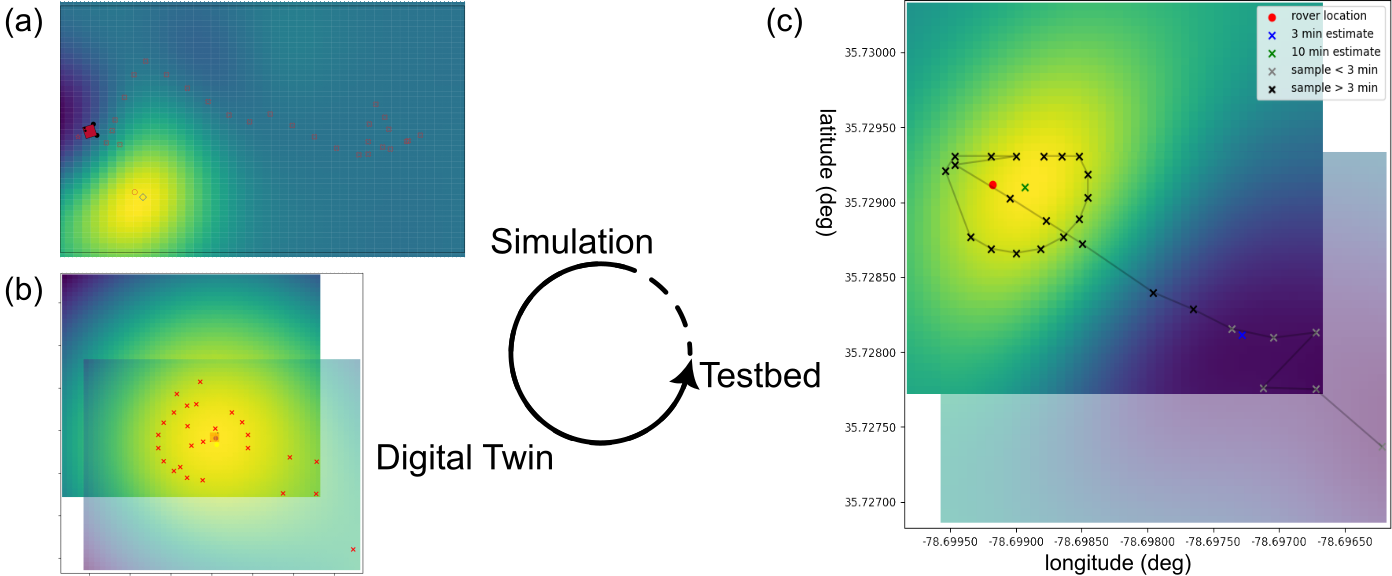} 
        \caption{Shows the progressive approach for creating a radio map estimate. (a) is a final simulation using Robotarium. (b) is a digital twin emulation test using AERPAW, and (c) is a reconstruction created using flight data logs from the second trial on the AERPAW testbed.} 
        \label{fig:uga} 
    \end{figure*}

Initially, the team modeled a smooth 2D path loss function in Robotarium~\cite{pickemRobotariumRemotelyAccessible2017}, a Python-based open-source robotic simulator, to analyze expected self-guided behavior. Using a 2D normal distribution for the basic path loss allowed control of the mean as a transmitter’s location and the covariance to model the decay. A simulation result is shown in Figure~\ref{fig:uga}a. The robotic platform tested various path loss models, signal noise levels, and agential behavior with different heuristic optimization acquisition functions~\cite{wangRecentAdvancesBayesian2023}. Predetermined routines of drone maneuvers before and after Bayesian optimization proved beneficial. The robot established a path loss direction with a few widely spaced samples, creating a reliable start that rapidly converged on the transmitter’s position. When repeatedly estimating a similar position, the robot could construct a circle of waypoints around that estimated position. This strategy reduced the risk of an incorrect radio map due to noise or anomalies in the radio samples and refined the final estimated position. By trialing different Bayesian optimization acquisition functions, the team chose to use the upper confidence bound function for final drone guidance.

After successful simulation testing in Robotarium, the team ensured the GP received valid and accurate path loss samples. This filtering step was crucial due to the limited testbed radio data. The receiver randomly encountered rapid signal fading, termed receiver ''dropout.'' It was important to exclude these dropout samples from the path loss estimate to avoid slowing down the drone’s estimation. The team set up a low-power laboratory testbed to gather receiver data. Tests confirmed that receiver movement increased the chance of encountering receiver dropout~\cite{matzFundamentalsTimeVaryingCommunication2011}. Various time-series filters were then trialed with the lab and AERPAW testbed data but could not remove the dropout. A final go-no-go filter was designed based on grouping the confidence level readings over short periods. The model excluded readings during high variance in the receiver's confidence level and discarded the averaged power as an outlier if it exceeded this quality-variance threshold.

With these preliminary steps using the open-source SDR and Robotarium platforms, the team developed a final algorithm to deploy on AERPAW. The UAV continually updated a grid of points predefined by latitude and longitude to represent the GP's radio map. The UAV used two maps: one for UAV guidance and another to estimate the UGV position. These maps correspond to the separated boundaries seen in Figure~\ref{fig:aerpaw_locations}, and a final digital twin result is shown in Figure~\ref{fig:uga}b. The UAV began its mission with a quality-variance threshold set using the existing testbed data. However, after takeoff and facing northwest, the UAV could exponentially increase the threshold until it accepted this first sample. After taking this first critical sample, the UAV went on to sample three additional waypoints before autonomously selecting the next waypoint using Bayesian optimization. The UGV radio map provided the three-minute and ten-minute estimates. Figure~\ref{fig:uga}c shows a final estimate of the second testbed run.

Throughout the effort to construct a rapid radio localization algorithm, the team carefully tuned the necessary parameters and initial settings of the GP kernel functions.
Success in the Robotarium simulation and AERPAW digital twin environment confirmed the algorithm’s task effectiveness within the designated time frame without violating any boundary constraints. Lab data collection also allowed for further enhancements of the final algorithm. Still, the team conservatively managed any parameter adjustments to account for discrepancies and limitations expected from these mock environments.

\section{Results} 
\label{sec:results}

\begin{figure}[!ht]
    \centering
    \includegraphics[width=0.93\linewidth]{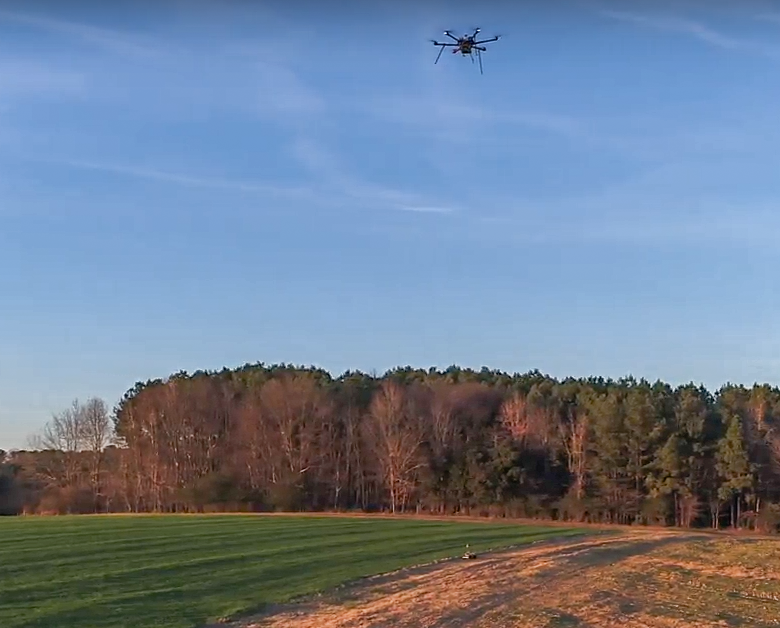}
    \caption{Snapshot taken during the competition with the UAV searching for the rover at Location 3.}
    \label{fig:duringExperiment}
\end{figure}

The emulation and testbed performances were completed by mid-December 2023. For the testbed run, all fifteen experiments (three positions for five finalist teams) were completed on the same day from approximately 10~am to 5~pm. Figure~\ref{fig:duringExperiment} shows an instance where the UAV was passing close to the rover's location while the rover was at the third hidden location.

Overall, each team demonstrated strong performances in specific runs, highlighting the necessity for various approaches and revealing distinct insights on a team's performance within specific circumstances. Additionally, broader trends and patterns emerge by examining the results across all runs for each team. The results are shown in Table \ref{tab:results}.

Within the emulator, NYU showed stellar performances in most individual runs, giving excellent final estimate consistency, and swept the win for both the three-minute and ten–minute estimates. 
 
\begin{table*}[!ht]
\centering
\caption{Simulation results and final testbed results for each team. The fast results correspond to each three-minute estimate,
\\while the final runs are given in ten minutes. Each run has the lowest error highlighted in blue.}
\begin{tabular}{|l|r|r|r|r|r|r|r|r|}
\hline
\rowcolor{gray!50}
Team         & Fast 1  & Fast 2  & Fast 3  & Fast Average & Final 1 & Final 2 & Final 3 & Final Average \\ \hline
\rowcolor{gray!25}
\multicolumn{9}{|c|}{Simulation Results (m)} \\ \hline
UGA & 71.7   & 265    & 174    & 170.23      & 52.8   & 17     & 19     & 29.6         
\\ \hline
UNT & \cellcolor{blue!25}10.9  & 15.7   & 70     & 32.2  & 15.5   & \cellcolor{blue!25}2.5    & 65     & 27.66  
\\ \hline
NYU & 22.7   & \cellcolor{blue!25}2.6    & \cellcolor{blue!25}61.1   & \textbf{28.8}        & \cellcolor{blue!25}1.97  & 2.7    & \cellcolor{blue!25}3.1    & \textbf{2.59}   
\\ \hline
\rowcolor{gray!25}
\multicolumn{9}{|c|}{Testbed Results (m)} \\ \hline
UGA & \cellcolor{blue!25}40.5   & 239.1  & 232.4  & 170.7       & 41.6   & \cellcolor{blue!25}27.8   & 130.4  & 66.6 \\ \hline
UNT & 88.9   & \cellcolor{blue!25}19.6   & \cellcolor{blue!25}77.4   & \textbf{62.0}        & \cellcolor{blue!25}17.6   & 48.3   & 77.3   & \textbf{47.8}         
\\ \hline
NYU & 99.8   & 143.4  & 79.8   & 107.7       & 64.3   & 67.3   & \cellcolor{blue!25}73.4   & 68.3
\\ \hline
\end{tabular}
\label{tab:results}
\end{table*}

The testbed results show much more variation among the team's results, highlighting the challenge and gap between the real-world testbed and a more controlled digital twin environment. The first run's three-minute result shows a relatively accurate result from UGA due to the short start routine before full autonomous control. For the final estimate, UNT improved its final estimate significantly, returning the best estimate within the entire testbed trials, within 17.6~m from the rover's true location. 
However, the second run gives distinctly different results. UNT gives the best three-minute estimate by a wide margin. UGA has the largest estimated improvement, showing a team-best accuracy of 27.8~m from the rover location.
The third run shows similar three-minute results from the UNT and NYU teams. However, the NYU team improved their estimate, giving the best final ten-minute estimate for the most challenging trial.

\section{Conclusions and Analysis} 
\label{sec:conclusions}
The UNT team secured a first-place award across both categories from the final testbed results. The UGA and NYU teams each earned second and third-place finishes in the two categories. The collaborative achievements of the AERPAW team and all participating schools mark the inaugural successful completion of the AERPAW Challenge, and the data collected is now accessible to the public. 

Anticipating this new data, efforts are already underway to integrate the data collected into improving the emulator to continue closing the digital twin loop. The heterogeneous results across the digital twin emulation and testbed clearly show a need for wireless community platforms that unlock the highest levels of wireless research. Mutual efforts like these provide training and data for a robust wireless future.
Please stay tuned for more exciting developments and similar competitive challenges.

\section*{Acknowledgments} 
The three participating teams (NYU, UNT, UGA) are grateful to AERPAW and all sponsors for the exciting and challenging educational experience.

\vfill 

\vfill 

\section{Biography Section} 
\vspace{-33pt}
\begin{IEEEbiographynophoto}{Paul Kudyba}
recently received his master's in electrical and computer engineering from The University of Georgia. He received his B.S. in 2015 from Southern Polytechnic State University, now Kennesaw State. He is currently seeking opportunities for collaboration in machine learning for industrial processes and Internet of Things.
\end{IEEEbiographynophoto}
\vspace{-33pt}
\begin{IEEEbiographynophoto}{Jaya Sravani Mandapaka}
is currently pursuing her Ph.D in Electrical Engineering at the University of North Texas. Her research is focused on UAS-UAS Communications- use case scenarios and message protocols. Jaya received her master's degree in computer science engineering from Arkansas State University in May 2021 and her B.S. in Electrical and Communication Engineering from Jawaharlal Nehru Technological University Hyderabad, India. Her research interests lie in communication Advancements of Advanced Air Mobility (AAM) vehicles.
\end{IEEEbiographynophoto}
\vspace{-33pt}
\begin{IEEEbiographynophoto}{Weijie Wang}
received a B.S. degree in Computer Engineering from New York University Tandon School of Engineering, New York, United States in 2023. He is working toward a M.S. degree at Columbia University, New York, United States. His research interests include UAV-assisted wireless communication, embedded systems, and Internet of Things.
\end{IEEEbiographynophoto}
\vspace{-33pt}
\begin{IEEEbiographynophoto}{Logan McCorkendale}
recently graduated with his Bachelor of Science in Electrical Engineering from the University of North Texas (UNT).  His research is focused on autonomous system design for the advancement of Advanced Air Mobility (AAM).
\end{IEEEbiographynophoto}
\vspace{-33pt}
\begin{IEEEbiographynophoto}{Zachary McCorkendale}
is a recent graduate from the Department of Electrical Engineering at the University of North Texas. His research interests include Autonomy and UAV communications for the Advancement of Air Mobility.
\end{IEEEbiographynophoto}
\vspace{-33pt}
\begin{IEEEbiographynophoto}{Mathias Feriew Kidane}
recently received his B.S. degree in Electrical Engineering from the University of North Texas.
\end{IEEEbiographynophoto}
\vspace{-33pt}
\begin{IEEEbiographynophoto}{Haijian Sun} 
is an Assistant Professor in the School of Electrical and Computer Engineering at The University of Georgia. He obtained his Ph.D. in the Department of Electrical and Computer Engineering from Utah State University, USA. His current research interests include vehicular communication, wireless communication for 5G and beyond, IoT communications, and optimization analysis.  Dr. Sun is a Member of the IEEE.
\end{IEEEbiographynophoto}
\vspace{-33pt}
\begin{IEEEbiographynophoto}{Eric Adams}
A longtime researcher at the University of Massachusetts Amherst ECE, he is the founder of 3 companies relating to remote sensing, edge-to-cloud computing, and cyberinfrastructure for small UAS and Advanced Air Mobility.
\end{IEEEbiographynophoto}
\vspace{-33pt}
\begin{IEEEbiographynophoto}{Kamesh Namuduri}
is a professor in the Department of Electrical Engineering at the University of North Texas. His research interests include  Autonomy and UAV communications.
\end{IEEEbiographynophoto}
\vspace{-33pt}
\begin{IEEEbiographynophoto}{Fraida Fund}
is a Research Assistant Professor in the Department of Electrical and Computer Engineering at the NYU Tandon School of Engineering.
She received her Ph.D. degree in Electrical Engineering from NYU Tandon School of Engineering. 
Her research interests include low latency wireless network protocols, economics of wireless networks, and design of open experimental platforms for research and education in communication networks. 
\end{IEEEbiographynophoto}
\vspace{-33pt}
\begin{IEEEbiographynophoto}{Mihail Sichitiu}
is a professor in the Department of Electrical Engineering at NC State University. His primary research interest is in Wireless Networking with an emphasis on multi-hop networking and wireless local area networks.
\end{IEEEbiographynophoto}
\vspace{-33pt}
\begin{IEEEbiographynophoto}{Ozgur Ozdemir}
is an Associate Research Professor at the Department of Electrical and Computer Engineering at NC State University. His research interests include mmWave channel sounding, SDRs, and UAV communications.
\end{IEEEbiographynophoto}


\begin{thebibliography}{1} 
\bibitem{kwonRFSignalSource2023a}
H.~Kwon and I.~Guvenc, ``{{RF Signal Source Search}} and {{Localization Using}} an {{Autonomous UAV}} with {{Predefined Waypoints}},'' in \emph{{{IEEE}} 97th {{Vehicular Technology Conference}} ({{VTC2023-Spring}})}, pp. 1--6, 2023.

\bibitem{AERPAW}
V.~Marojevic  \emph{et al.}, ``Advanced wireless for unmanned aerial systems: 5g standardization, research challenges, and aerpaw architecture,'' \emph{IEEE Vehicular Technology Magazine}, vol.~15, no.~2, pp. 22--30, 2020.

\bibitem{rasmussenGaussianProcessesMachine2006}
C. E. Rasmussen and C. K. I. Williams, Gaussian processes for machine learning. in \emph{Adaptive computation and machine learning. Cambridge, Mass: MIT Press}, 2006. 

\bibitem{bayesopt}
F.~Nogueira, ``{Bayesian Optimization}: Open source constrained global optimization tool for {Python},'' accessed: 2024-05-30. [Online]. Available: \url{https://github.com/bayesian-optimization/BayesianOptimization}.

\bibitem{PyturfPyturf2024}
``Pyturf/pyturf,'' pyturf, accessed: 2024-5-30. [Online]. Available: \url{https://github.com/pyturf/pyturf}.

\bibitem{wilsonEfficientlySamplingFunctions2020}
J.~Wilson, \emph{et al.}, “Efficiently sampling functions from Gaussian process posteriors,” in \emph{Proceedings of the 37th International Conference on Machine Learning},   pp. 10292–10302, Nov. 2020.

\bibitem{santosMultirobotLearningCoverage2021}
M.~Santos \emph{et al.}, ``Multi-robot {{Learning}} and {{Coverage}} of {{Unknown Spatial Fields}},'' in \emph{{{International Symposium}} on {{Multi-Robot}} and {{Multi-Agent Systems}}}, pp. 137--145, 2021.

\bibitem{finkOnlineMethodsRadio2010}
J.~Fink and V.~Kumar, ``Online methods for radio signal mapping with mobile robots,'' in \emph{{{IEEE International Conference}} on {{Robotics}} and {{Automation}}}, pp. 1940--1945, 2010. 

\bibitem{shresthaRadioMapEstimation2023}
R.~Shrestha \emph{et al.}, ``Radio {{Map Estimation}} in the {{Real-World}}: {{Empirical Validation}} and {{Analysis}},'' in \emph{{{IEEE Conference}} on {{Antenna Measurements}} and {{Applications}}}, pp. 169--174, 2023.

\bibitem{duvenaudAutomaticModelConstruction}
D.~K. Duvenaud, ``Automatic {{Model Construction}} with {{Gaussian Processes}},'' accessed: 2023-05-25. [Online]. Available: \url{https://www.cs.toronto.edu/~duvenaud/thesis.pdf}

\bibitem{matthewsGaussianProcessBehaviour2018a}
Matthews, Hron, Rowland, Turner, and Ghahramani, “Gaussian Process Behaviour in Wide Deep Neural Networks,” in \emph{ International Conference on Learning Representations}, Feb. 2018. 

\bibitem{deisenrothGaussianProcesses}
M.~Deisenroth, ``Gaussian {{Processes}},'' accessed: 2024-05-30. [Online]. Available: \url{https://www.deisenroth.cc/teaching/2019-20/linear-regression-aims/lecture_gaussian_processes.pdf}

\bibitem{wangRecentAdvancesBayesian2023}
X. Wang \emph{et al.}, “Recent Advances in Bayesian Optimization,” ACM Comput. Surv., vol. 55, no. 13s, pp. 287:1-287:36, Jul. 2023.

\bibitem{pickemRobotariumRemotelyAccessible2017}
D.~Pickem \emph{et al.}, ``The {{Robotarium}}: {{A}} remotely accessible swarm robotics research testbed,'' in \emph{{{IEEE International Conference}} on {{Robotics}} and {{Automation}}}.\hskip 1em plus 0.5em minus 0.4em\relax IEEE, pp. 1699--1706, 2017. 


\bibitem{matzFundamentalsTimeVaryingCommunication2011}
G.~Matz and F.~Hlawatsch, ``Fundamentals of {{Time-Varying Communication Channels}},'' in \emph{Wireless {{Communications Over Rapidly Time-Varying Channels}}}.\hskip 1em plus 0.5em minus 0.4em\relax Elsevier, pp. 1--63, 2021. 


\bibliographystyle{IEEEtran}
\end{thebibliography}
\end{document}